\documentclass[aps,amsmath,amssymb,twocolumn,showpacs,prx, longbibliography]{revtex4-1}

\usepackage{graphicx} 
\usepackage{dcolumn}
\usepackage{bm}
\usepackage{amsmath}
\usepackage{amssymb}
\usepackage{color}
\usepackage{bbold}
\usepackage{bm}

\renewcommand{\vec}[1]{{\boldsymbol #1}}
\newcommand{\ph}{\text{ph}}

\usepackage{natbib}

\newcommand{\tr}{\text{Tr}}

\newcommand{\llangle}[1][]{\savebox{\@brx}{\(\m@th{#1\langle}\)}%
  \mathopen{\copy\@brx\kern-0.5\wd\@brx\usebox{\@brx}}}
\newcommand{\rrangle}[1][]{\savebox{\@brx}{\(\m@th{#1\rangle}\)}%
  \mathclose{\copy\@brx\kern-0.5\wd\@brx\usebox{\@brx}}}
  \newcommand{\kxy}{\kappa_{xy}}

\begin{document}
\title{Approximately quantized thermal Hall effect of chiral liquids coupled to phonons}
\author{Yuval Vinkler-Aviv}
\author{Achim Rosch}
\affiliation{Institute for Theoretical Physics, University of Cologne, 50937 Cologne, Germany}

\begin{abstract}

The recent observation of a half-integer quantized thermal Hall effect in $\alpha$-RuCl$_3$ is interpreted as a unique signature of a chiral spin liquid with a Majorana edge mode. A similar quantized thermal Hall effect is expected in chiral topological superconductors. The unavoidable presence of gapless acoustic phonons, however, implies that, in contrast to the quantized electrical conductivity, the thermal Hall conductivity $\kxy$ is never exactly quantized in real materials. Here, we investigate how phonons affect the quantization of the thermal conductivity focusing on the edge theory. As an example we consider a Kitaev spin liquid gapped by an external magnetic field coupled to acoustic phonons.
The coupling to phonons destroys the ballistic thermal transport of the edge mode completely, as energy can leak into the bulk, thus drastically modifying the edge-picture of the thermal Hall effect. Nevertheless, the thermal Hall conductivity remains approximately quantized and we argue, that the coupling to phonons to the edge mode is a necessary condition for the observation of the quantized thermal Hall effect. The strength of this edge coupling does, however, not affect the conductivity. We argue that for sufficiently clean systems the leading correction to the quantized thermal Hall effect, $\Delta \kappa_{xy}/T \sim \text{sign(B)} \, T^2$, arises from a intrinsic anomalous Hall effect of the acoustic phonons due to Berry phases imprinted by the chiral (spin) liquid in the bulk. This correction depends on the sign but not the amplitude of the external magnetic field.
\end{abstract}

\maketitle
\section{introduction}

Quantum Hall effects are prime examples showing how topological properties of matter
lead to a precise quantization of physical observables. The integer and fractional quantum Hall effects 
are characterized by quantized Hall conductivity $\sigma_{xy}=\frac{n}{m} \frac{e^2}{2 \pi \hbar}$ with integer $n,m$ and by a vanishing longitudinal conductivtiy $\sigma_{xx}=0$. A quantum Hall effect in the charge sector is usually also associated with a thermal quantum Hall effect. The latter can, however, also occur separately. Both for topological superconductors and insulators \cite{Read_2000,  Wang_2011, Stone_2012} and for certain chiral spin liquids (e.g., described by the Kitaev model \cite{Kitaev_2006}) 
a quantized thermal Hall effect, $\frac{1}{T}\kappa_{xy}=\frac{n}{m}\frac{\pi k_B^2}{6\hbar}$, has been predicted in close analogy to the (anomalous) Hall effect in two-dimensional conductors. 

Recently, a spectacular experiment by Kasahara {\it et al.} \cite{Kasahara_2018} reported the discovery of 
a half-integer quantized thermal Hall effect in $\alpha$-RuCl{\textsubscript 3}. This  material is believed to be approximately described by a Kitaev model on a honeycomb lattice~\cite{Nasu_2014, Plumb_2014, Kim_2015, Johnson_2015, Sears_2015, Sandilands_2015, Sandilands_2016a, Sandilands_2016b, Banerjee_2016, Zhou_2016, Koitzch_2016, Sinn_2016, Ziatdinov_2016, Cao_2016, Kim_2016, Yadav_2016, Chaloupka_2016, Nasu_2016, Trebst_2017, Hirobe_2017, Hickey_2018, Banerjee_2018} While it orders magnetically at zero magnetic field, Kasahara {\it et al.} identified a range of fields where the magnetic order is apparently suppressed and a gapped spin liquid characterized by a quantized thermal Hall effect emerges. Such a phase has been predicted by Kitaev \cite{Kitaev_2006} in the idealized honeycomb Kitaev model. The two-dimensional honeycomb Kitaev model is an example of a frustrated spin liquid which can accurately be described by Majorana fermions coupled to a $\mathbb{Z}_2$ gauge field. In the presence of a magnetic field a gap opens up in the bulk and a chiral edge modes of Majorana fermions arise at the boundaries of the sample. The latter carries heat and leads to a quantized thermal Hall effect, $\frac{1}{T}\kappa_{xy}=\frac{1}{2}\frac{\pi k_B^2}{6\hbar}$, precisely as observed experimentally.

There is a major difference between the electric and the thermal Hall effect when realized in real material: in all solids phonons provide a thermal conduction mechanism, leading to a finite heat conductivity $\kappa_{xx}$ typically with $\kappa_{xx} \gg \kappa_{xy}$.  In contrast, a hallmark of the quantized electrical Hall response is a vanishing bulk conductivity. Theoretical models predicting a thermal quantum Hall effect usually ignore the presence of phonons (thus assuming or concluding that 
$\kappa_{xx}=0$) and sometimes it is argued that the coupling of phonons will prohibit the observation of a quantized thermal Hall effect.

There are two complementary ways to understand quantum Hall effects: the bulk and the edge perspective. From the first point of view, Hall effects are bulk phenomena. The thermal conductivity can be calculated, e.g., from an appropriate Kubo formula in large system with periodic boundary conditions. An alternative and more intuitive point of view explains quantum Hall effects by the presence of chiral conducting channels that exist at the edge of the sample. The presence of the edge channels is enforced and protected by the (insulating) bulk theory.
The resulting bulk-edge correspondence is a manifestation of the holographic principle and intimately connected to topological field theories and the physics of anomalies. Charge anomalies reflect that a change of charge in the bulk of the system (e.g. when the magnetic field is changed) can only occur 
at the gapless edge of a system. The change of magnetic field thereby induces an electric field which pumps charges into chiral edge modes. For the thermal Hall effect similar arguments exist where the electromagnetic field is replaced by a (fictitious) gravitational field~\cite{Luttinger_1964,Witten_1984, Read_2000, Cappelli_2002, Schnyder_2008, Ryu_2010, Stone_2012, Ryu_2012, Nomura_2012, Gromov_2015}, {\it i.e.} a change of the metric, which leads to pumping of energy instead of charge. When discussing anomalies one usually assumes a gapped bulk, and it is a priori not fully clear how the presence of gapless phonons will affect the result and the quantization of the Hall effect. 

Edge theories of the Hall effect use the fact that chiral modes can transport charge or energy without dissipation. The latter is, however, not valid in the presence of phonons which destroy ballistic edge transport. Heat leaks out from the edge into the bulk. Therefore recent measurements by the Weizmann group \cite{Banerjee_2017a,Banerjee_2017b} of the thermal transport of (fractional) quantum Hall edge channels had to use a refined setup which involved {\em mesoscopic} measurements of the temperature of edge channels using noise spectroscopy.
With this setup the authors could minimize the leakage of heat due to phonons. In constrast, the experiments in $\alpha$-RuCl$_3$ by Kasahara {\it et al.} \cite{Kasahara_2018} used macroscopic contacts and operated in the opposite regime where edge states and phonons are expected to equilibrate. We will show that, counter-intuitively, the phonon leakage of heat actually helps to observe an approximately quantized thermal Hall effect. We will discuss both two dimensional models and  three dimensional systems described, e.g., by weakly coupled layers of Kitaev models in the presence of three-dimensional phonons.

It is obvious that in the presence of phonons a true quantization of the thermal Hall effect is not possible. Here `true quantization' is defined as an effect which becomes more and more quantized {\em with exponential precision} when the temperature is lowered and the size of the system is increased. Any insulator in a magnetic field (or in a ferromagnetic state) is characterized by a finite thermal Hall effect at finite temperatures, which is not exponentially suppressed with temperature but expected to vanish with a power law of $T$. Nevertheless, the observed phonon thermal Hall effects of insulator are typically tiny. Very few experiments \cite{Sugii_2017} exist, the first one from 2005 \cite{Strohm_2005} reports a Hall angle of only  $10^{-4}$ rad per Tesla of applied field.  Several mechanism can explain phonon Hall effects. In the literature
a Raman-type interaction between phonons and large spins~\cite{Sheng_2006,Wang_2009}, effects of Berry curvature of phonon bands~\cite{Zhang_2010,Qin_2012}, and a resonant skew scattering of phonons~\cite{Mori_2014} has been discussed.

In the following, we will discuss the approximate quantization of the thermal Hall effect from the viewpoint of the edge theory and develop a theory of an anomalous thermal Hall effect based on the coupling of phonons to the Kitaev model.

\section{Thermal Hall effect}
\subsection{Quantized ballistic Hall effect in the absence of phonons}\label{ballisticSecNoPhonons}

\begin{figure}[t]
\begin{tabular}{cccc}
(a) & \includegraphics[width=0.2\textwidth]{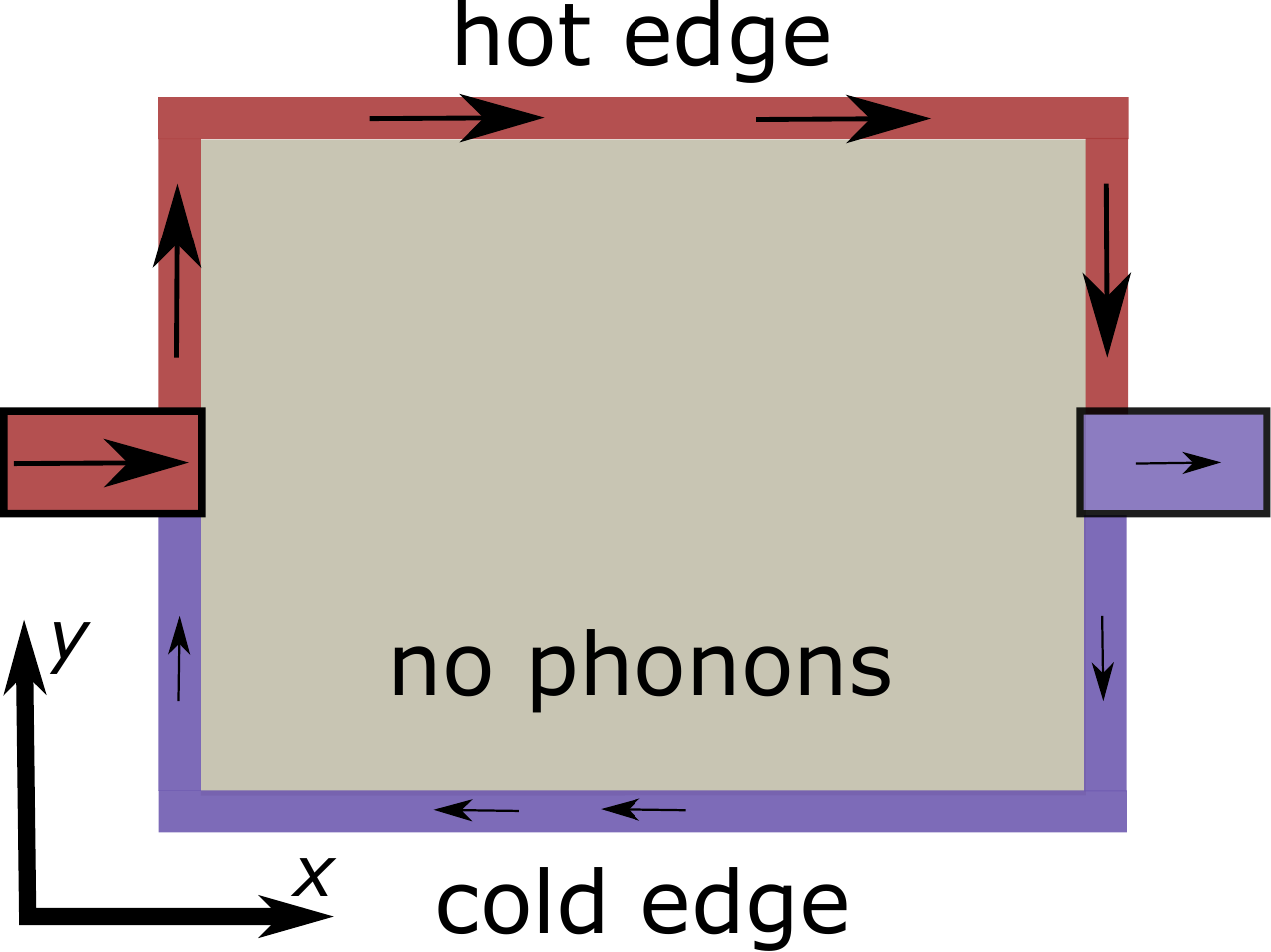}
& (b) &
\includegraphics[width=0.2\textwidth]{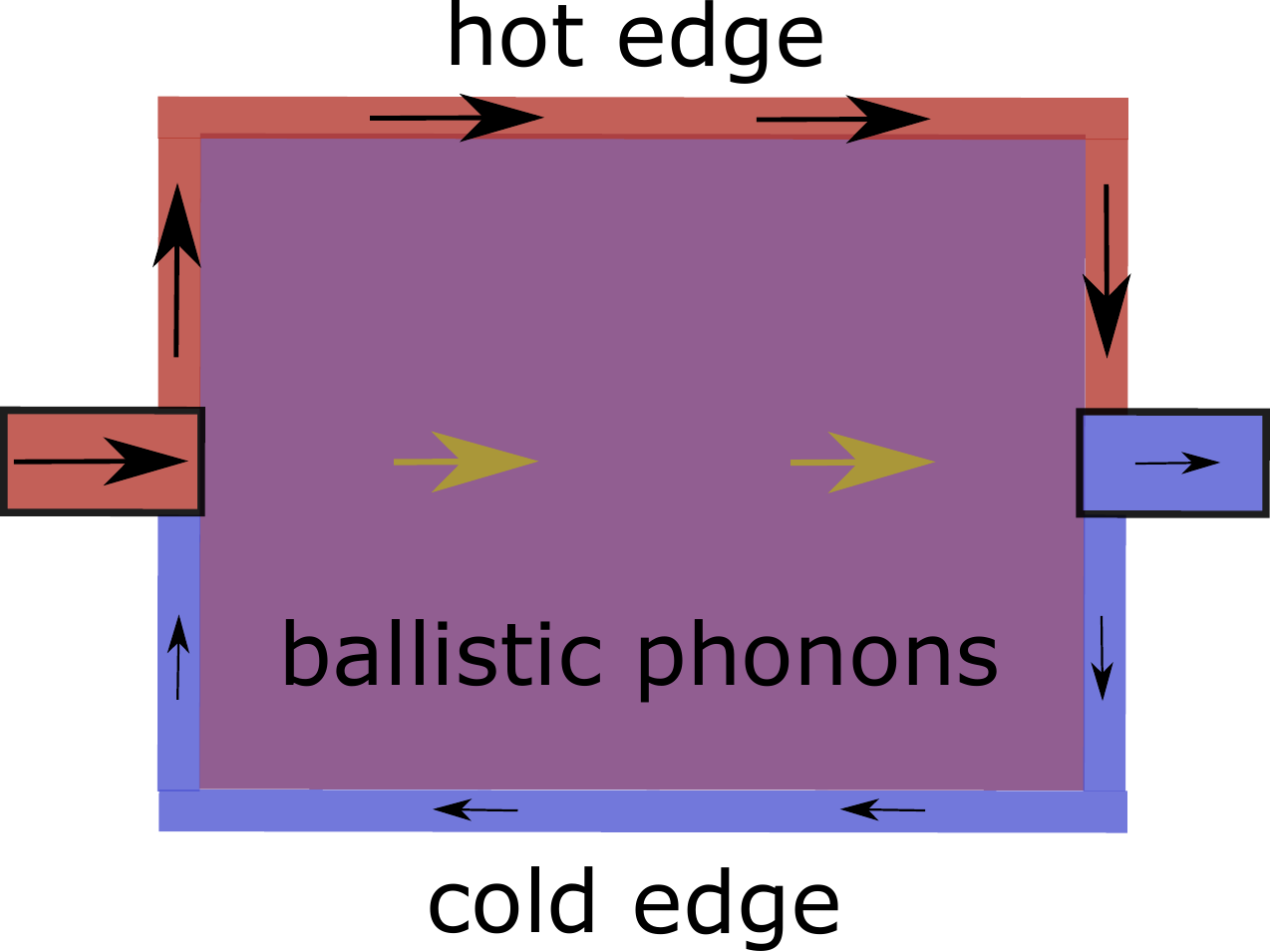} \\ \\
(c) & \includegraphics[width=0.2\textwidth]{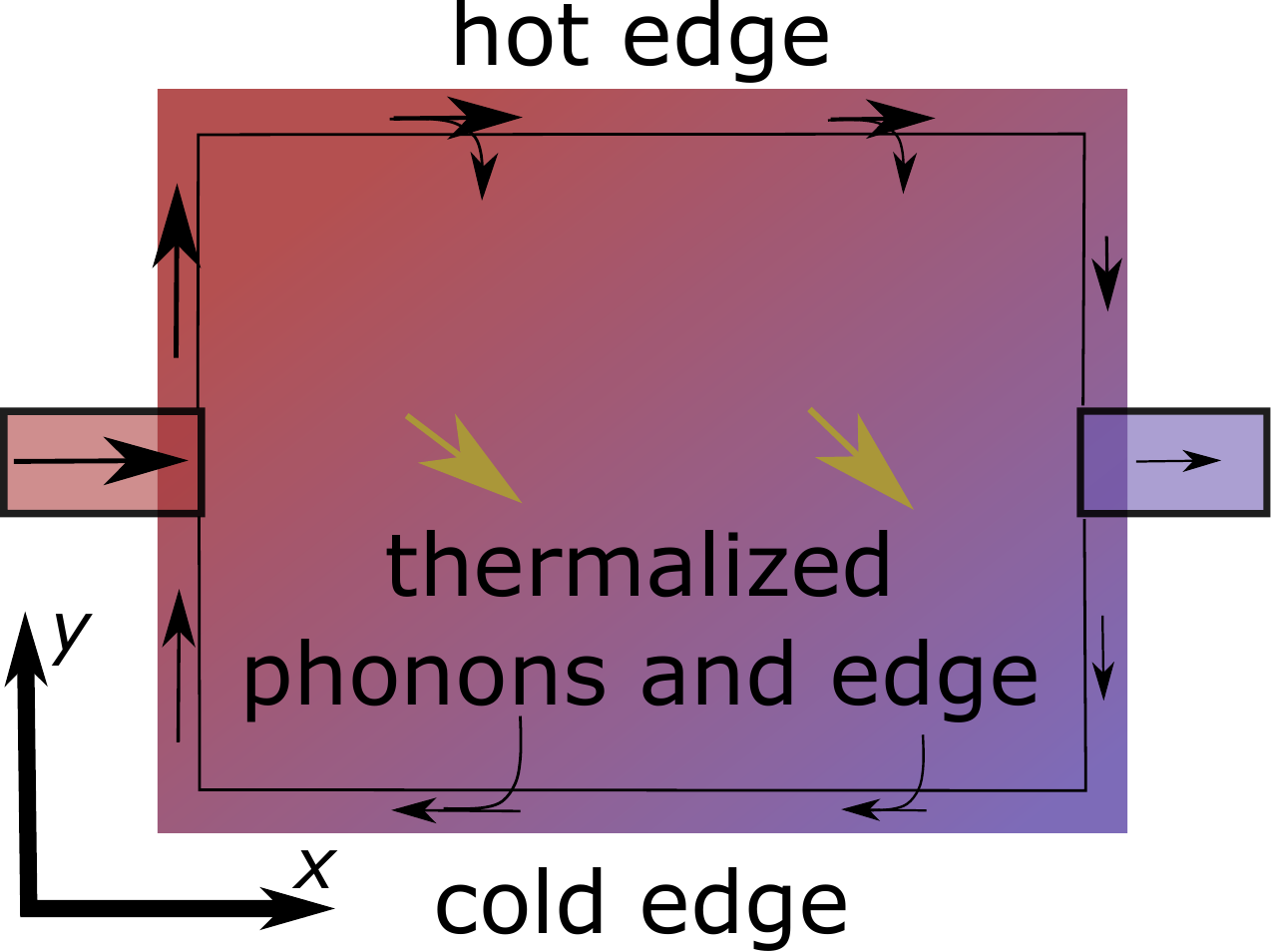}
& (d) & \includegraphics[width=0.21\textwidth]{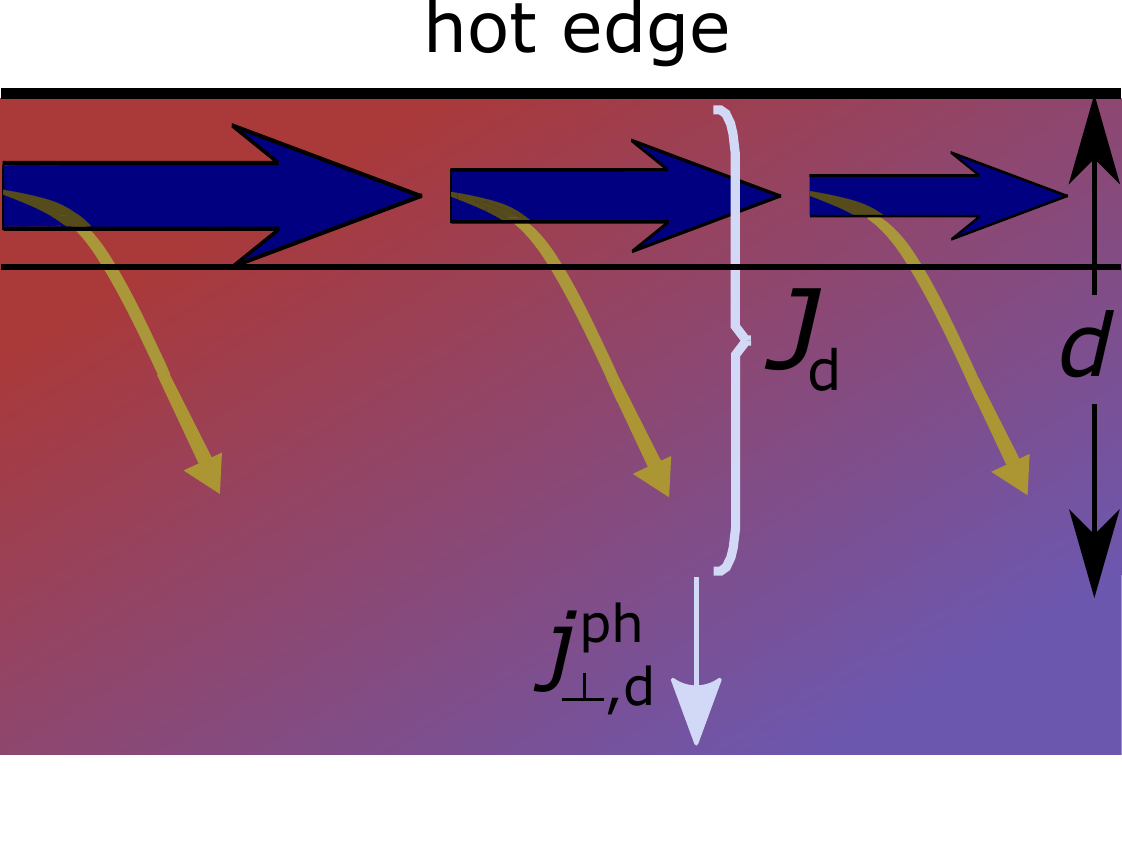}
\end{tabular}
\caption{(color online)
Sketch of heat currents (arrows) and local temperatures (color scale) in the presence of chiral edge channels.  In the absence of phonons, panel a), heat is transported ballistically by the edge channel resulting in a quantization of $\kappa_{xy}/T$. Panel b) sketches a regime where the sample size is smaller than the phonon-edge scattering length such that the edge states and phonons have different temperatures. No quantization of $\kappa_{xy}$ is expected in this case. In the opposite limit, panel c), the edge channel and the phonons have the same local temperatures. The temperature gradient imprinted by the phonons onto the edge channel implies that a quantized heat current is injected from the edge mode into the bulk phonons as sketched in panel d). This leads to quantization of $\kappa_{xy}$ with high precision.}
\label{fig:setup}
\end{figure}
We begin our discussion by recalling how the thermal thermal Hall effect is described in terms of a chiral edge mode in the {\em absence} of phonons, when the bulk is a thermal insulator. In this case thermal transport is only possible along a chiral edge mode. Fig. \ref{fig:setup}a shows a sketch of such a situation. The chiral edge mode carries a finite heat current which is a function of temperature, $J_e(T)$. As in the absence of bulk modes the energy density at the edge $E_e(\vec r,t)$ is conserved, $\partial_t E+\nabla_\| J_e=0$, the heat current is constant in the stationary state, $\nabla_\| J_e=0$. Here $\nabla_\|$ is a spatial derivative parallel to the edge of the system. Therefore the temperatures along the top and bottom side of the sample, $T_{\rm top}$ and $T_{\rm bot}$, are constant and the total energy current through the system is given by $J_T=J_e(T_{\rm Top})-J_e(T_{\rm Bot})$. One then finds in the ballistic case and in absence of phonons for small temperature difference $T_{\rm top}-T_{\rm bot}$
\begin{eqnarray}
	J_T = \frac{d J_e(T)}{d T} (T_{\rm top}-T_{\rm bot})
\end{eqnarray}
and therefore
\begin{eqnarray}
\kappa_{xy}=\rho^{th}_{yx}= \frac{d J_e(T)}{d T}, \qquad \kappa_{xx}=\rho^{th}_{xx}=0 \label{ballistic}
\end{eqnarray}
Here the thermal resistivity tensor is defined as the inverse of the thermal conductivity tensor, $\hat \rho^{th}=\hat \kappa^{-1}$.
To simplify notations, we assume in the following a system with $90^{\circ}$ rotation invariance such that $\kappa_{xx}=\kappa_{yy}$ and
$\kappa_{xy}=-\kappa_{yx}$. In Eq.~\eqref{ballistic} we used that there is no temperature drop parallel to the direction of the edge channels implying that $\rho^{th}_{xx}=0$ and therefore also
   $\kappa_{xx}=\frac{\rho^{th}_{xx}}{(\rho^{th}_{xx})^2+(\rho^{th}_{xy})^2} =0$. 
   
The formula Eq.~(\ref{ballistic}) is valid for a 2d system. Its generalization to three dimensions is obtained by dividing the result by the thickness $L_z$ of the system
   \begin{eqnarray}
\kappa^{3d}_{xy}= \frac{1}{L_z} \frac{d J_e(T)}{d T} \label{ballistic3d}
\end{eqnarray}   

It is instructive to calculate the thermal derivative of the heat current for non-interacting electrons. For a single chiral fermionic channel with arbitrary dispersion $\epsilon_{k_x}$ one finds
\begin{align}
\frac{d J_e(T)}{d T}&=&\int \epsilon_{k_x} \frac{d f(\epsilon_k)}{d T} \frac{d \epsilon_{k_x}}{d \hbar k_x} \frac{d k_x}{2 \pi} = - \int_{\epsilon_{\rm min}}^{\epsilon_{\rm max}} \frac{\epsilon^2  f'(\epsilon)}{2 \pi \hbar k_B T} \, d\epsilon 
\end{align}
The remaining integral can be done analytically and one obtains the textbook result~\cite{Cappelli_2002}
\begin{align}\label{quantized}
\frac{1}{T} \frac{d J_e(T)}{d T}&=& (c_R - c_L) \frac{\pi k_B^2}{6 \hbar } +O\left(\exp\!\left[-\frac{\epsilon_{\rm min/max}}{k_B T}\right]\right)
\end{align}
We have added the prefactor $c_R-c_L$ which is equal to $1$ for the single chiral fermionic channel considered above. In general, it encodes the difference of the conformal charges of chiral conformal field theories describing the low-energy properties of the edge theory.
It is important to note that correction to the quantized result Eq.~\eqref{quantized} are exponentially suppressed when the temperature is smaller than the bulk gap which enters as the cutoff $\epsilon_{\rm min/max}$ of the edge theory. The quantization with exponential precision holds even when the edge theory is interacting and disordered which we will prove explicitly below, see Sec.~\ref{edgeAnomaly}.
For a single chiral Majorana channel one obtains half of the result of the fermionic channel, $c_R=1/2$, $c_L=0$ resulting in 
\begin{align}
\frac{\kappa_{xy}}{T}=\frac{1}{2} \frac{\pi k_B^2}{6 \hbar}
\end{align}
with exponential precision in the {\em absence} of phonons.

In a layered three dimensional system (e.g. a stack of weakly coupled Kitaev models in a perpendicular magnetic field) the number of edge channels $c_R-c_L$ is proportional to the number of layers $N_L$, resulting in a quantization of $
d_z\,\kappa_{xy}^{3d}$, where $d_z=L_z/N_L$ is the distance of layers, see Eq.~\eqref{ballistic3d}.

The quantization of $\frac{1}{T} \frac{d J_e(T)}{d T}$ and the corresponding quantization of $\kappa_{xy}/T$  for a system with a bulk gap strongly suggests a topological origin for this effect. As discussed in the introduction, it has been shown that this is indeed the case. The topology of the bulk theory is directly connected to a gravitational bulk and corresponding edge anomaly which explains the robustness of the result for systems with a bulk gap ({\it i.e.} without phonons) and the fact that corrections are exponentially small in the ration of bulk gap and temperature.

\subsection{Non-quantized ballistic Hall effect in the presence of phonons}\label{nonquantizedSec}
We now consider a system with (acoustic) phonons, where however, the size  of the sample $L$ is much {\em smaller} than a characteristic scattering length of chiral edge-excitations and phonons, $\ell_{e}^{\ph}\gg L$.
Note that the mean-free path of phonons grows rapidly with decreasing temperature
and therefore the ballistic heat transport by phonons, i.e., energy transport limited only by surface scattering of phonons, is routinely observed in insulators at very low temperatures even for macroscopically large samples \cite{Casimir_1938}.
In Fig.~\ref{fig:setup}b, we have sketched a situation where phonons are present 
but do not couple to the edge mode. The sketch assumes that the phonons are completely ballistic so that there is no temperature gradient inside the sample.

In the ballistic limit there are two independent channels to transport energy: the chiral edge mode and the bulk phonons. Correspondingly, the total heat current splits into an edge and a phonon contribution
\begin{eqnarray}
J_T&=&J_e+J_{\ph}=\left(\frac{d J_e(T)}{d T} +G^{th}_\ph \right) (T_{\rm top}-T_{\rm bot})
\end{eqnarray}
where $G^{th}_\ph$ is the ballistic conductance of the phonon system which depends in a non-universal way
on the properties of the contact.

In this situation obviously no quantization of the thermal Hall effect can be expected. The first problem is that the phonons and the edge mode are not in thermal equilibrium, a local temperature of the full system cannot be defined and thermometers attached to the same position will give different readings depending on whether they couple more to the phonons or to the edge mode. Two thermal contacts in transverse direction will therefore measure a temperature difference $\Delta T_{ex}=\alpha (T_{\rm top}-T_{\rm bot})$, where $0\le \alpha \le 1$ is a non-universal factor which depends on the precise location and coupling of the thermal contact.
Deep in the ballistic regime, the longitudinal energy gradient vanishes, $\rho^{th}_{xx}=0$. 
One then finds in the ballistic regime that 
\begin{align}
\kappa_{xy}=\frac{\rho^{th}_{xy}}{(\rho^{th}_{xx})^2+(\rho^{th}_{xy})^2}=\frac{J_T}{\Delta T_{ex}}=\frac{1}{\alpha}\left( \frac{d J_e(T)}{d T} +G^{th}_\ph \right)
\end{align}
which is non-universal and depends on the details of the contacts injecting the energy and used to measure the temperature drop. Depending on the contacts, it is possible to obtain values of $\kappa_{xy}$ which can be much larger or much smaller than the quantized value. As discussed in the introduction, a recent experiment \cite{Banerjee_2017a,Banerjee_2017b} managed to measure a quantized thermal Hall effect of an electronic fractional Hall effect by carefully designing a temperature probe (noise spectroscopy of electrons) with $\alpha\approx 1$ and by working in a regime where  $\frac{d J_e(T)}{d T} \gg G^{th}_\ph$. 

For generic thermal contacts, however, we conclude that in the ballistic limit phonons destroy any quantization of thermal Hall effects.

\subsection{Approximately quantized Hall effect due to phonon coupling}\label{SecThermal}

Finally, we consider a system in the thermodynamic limit where the scattering length of phonons with the edge mode and the  phonon-phonon scattering length are much smaller than the system size, $\ell_{\ph}^e,\ell_{\ph}^{\ph}\ll L$. This is the relevant limit for practically all bulk experiments (with the exception of very low temperatures when phonon transport can become ballistic as discussed above). In this limit, local temperatures and therefore also thermal resistivities are well defined. To avoid spurious effects of contacts, the standard procedure to measure the thermal resistivity tensor is to heat the system on one side ({\it e.g.} electrically) with a known power. Therefore the total steady-state heat current $J_T$ is known. Thermometers aligned parallel and perpendicular to $J_T$ are then used to measure the temperature gradients in parallel and perpendicular directions, which allows to deduce the thermal resistivity tensor and by inversion the thermal conductivity tensor. The main experimental error in such a setup is usually that the position of the thermometers may not be known with sufficient precision.

The coupling to phonons implies that ballistic edge channels do not exist any more. Energy can and will leak out of the edge channels into the bulk. The edge properties are therefore fundamentally different compared to the ballistic case without phonons. Also $\kappa_{xx}$ arising from bulk photons will be finite and typically much larger than $\kappa_{xy}$. 

Nevertheless, there is a convincing line of argument which suggests that $\kappa_{xy}$ (but not $\rho^{th}_{xy}$) should remain approximately quantized. Ultimately, we expect that the conductivity tensor is a {\em bulk} quantity which can be calculated from a Drude formula for the total heat current density $\vec j_H$ arising from three contributions, $\vec j_H=\vec j_s + \vec j_{\ph}+\vec j_{s,ph}$: the heat current of the (gapped) bulk theory  of the electronic or spin degrees of freedom $\vec j_s$, the heat current of the phonons $j_{\ph}$ and the heat current arising from the coupling between the two, see Sec.~\ref{sec:bulk}. If one assumes that the phonons and the mixed terms give only small contributions, it is plausible that essentially only the quantized response of the gapped topological phase is measured.


A main goal of our paper is to consider this question not from the viewpoint of the bulk theory but from the viewpoint of an edge theory coupled to phonons. We exploit the fact that all gapped degrees of freedom can be integrated out and it has to be possible to reformulate the theory only in terms of the edge channel and bulk phonons {\em without} using properties of gapped bulk modes.
Our goal is to clarify how the coupling to phonons can {\em restore} an approximate quantization of $\kappa_{xy}$ in a theory where the edge channels are not ballistic and where $\kappa_{xx}$ is large.

The starting point of our analysis is the continuity equation for the thermal edge current. We consider a strip of width $d$ along the edge of the sample and define the total energy density $E_d(x)=\int_0^d d y \, e(x,y)$ as function of the coordinate $x$ parallel to the edge by integrating over the strip. Similarly, the heat current
$J_d(x)=\int_0^d d y \, j_H^x(x,y)$ is defined as an integral of the total heat current density in $x$ direction. Note that both $e$ and $j^x_H$ include contributions from the edge modes, phonons and their interactions. We use a two-dimensional notation here but all arguments can easily be generalized to three dimensional system where the edge is a planar system.

In this case the continuity equation reads
\begin{align}\label{continuity}
\partial_t E_d + \nabla_x J_d = j^{\ph}_{\perp,d}
\end{align}
Here $j^{\ph}_{\perp,d}(x)$ is the heat current in the direction perpendicular to the edge at coordinates $(x,d)$, see Fig.~\ref{fig:setup}d. We choose $d$ to be sufficiently far away from the edge so that $j^{\ph}_{\perp,d}$ is purely carried by bulk phonons with no contribution arising from edge states or the coupling of the edge states to the phonons.

We will now use that in the linear response limit there is a local thermal equilibrium of the edge mode and the bulk phonons due to their coupling. Furthermore,
due to the presence of the chiral edge state, $J_d=J_d(T)$ is finite at finite temperatures.
In the steady state we have $\partial_t E_d=0$, therefore the edges pump or absorb energy current into the bulk of the system.
\begin{align}
j^{\ph}_{\perp,d} = \nabla_x J_d = \frac{d J_d}{d T} \nabla_x T  
\label{jperp}
\end{align}
The first equality in Eq.~\eqref{jperp} follows from the continuity equation at steady-state. For the second equality we
expand $J_d$ in powers of $\nabla T$, $J_d(x)=J_d(T(x)) + \kappa_e \nabla_x T(x)+O((\nabla T)^2)$, where the first term is the heat current in thermal equilibrium. The contribution from the second term to Eq.~\eqref{jperp}, $\nabla_x (\kappa_e \vec \nabla T)=\frac{d \kappa_e}{d T} (\nabla_x T)^2 + \kappa_e \nabla_x^2 T$, vanishes in the linear response, large $L$ limit. Thus only the first one, $\nabla_x J_d(T(x))=\frac{dJ_d}{dT} \nabla_x T$, needs to be kept.

Eq.~\eqref{jperp} is the direct manifestation of a thermal edge anomaly: a gradient in temperatures pumps energy into the edge of the system which is then transported away by phonons. We will show in Sec.~\ref{edgeAnomaly} that $\frac{d J_d}{d T}$
can be identified with the edge anomaly $\frac{d J_e}{d T}$ from Eq.~\eqref{quantized} in the absence of phonons.

At the next step, we investigate what pattern of thermal gradient emerges due to the injection of the heat from the edge channel into the phonon system. The total heat current in $x$ direction has two contributions, one from the bulk phonons and a further contribution arising from the
temperature difference $L_y \nabla_y T$ of the edge modes (as discussed in Sec.~\ref{ballisticSecNoPhonons})   
\begin{eqnarray}\label{jT}
\frac{J_T}{L_y} &=& j_\|^\ph+ \frac{d J_d}{d T}  \nabla_y T 
\end{eqnarray}

The temperature gradients measured experimentally are obtained from the energy current densities $j_\|^\ph$ and $j_\perp^\ph$ and the bulk conductivity matrix of the phonons
\begin{eqnarray}
{\hat \kappa}^\ph  \cdot
\left(\begin{array}{c}
\nabla_x T \\ \nabla_y  T
\end{array}
\right) = 
\left(
\begin{array}{c}
j_\|^\ph  \\ j_\perp^\ph
\end{array}
\right) = \left(
\begin{array}{c}
\frac{J_T}{L_y} - \frac{d J_d}{d T} \nabla_y T \\ \frac{d J_d}{d T} \nabla_x T
\end{array}
\right) 
\end{eqnarray}
where we used Eq.~\eqref{jperp}  and \eqref{jT}.
The interpretation of the experiment assumes that there is no current flowing in the perpendicular direction and that the current density in parallel direction is given by $J_T/L_y$. Therefore, $\nabla_x T=\rho^{th}_{xx} J_T/L_y$ and $\nabla_y T=\rho^{th}_{xy} J_T/L_y$. By inverting the $\hat \rho^\ph$ matrix we find 
\begin{eqnarray}\label{kappaAll}
\kappa_{xy}&=&\frac{d J_d}{d T} + \kappa_{xy}^\ph  \nonumber\\
\kappa_{xx}&=&\kappa_{xx}^\ph
\end{eqnarray}
This result, derived from the edge theory, has a straightforward interpretation: the heat conductivity of the low-energy bulk phonons has to be added to the 
conductivity arising from the edge channel -- as expected from the bulk-boundary correspondence.

This simple addition rule -- while highly plausible -- is a non-trivial result because usually (heat) conductivites are {\em not} additive. When one writes the total heat current, $\vec j=\vec j_\ph+\vec j_s+\vec j_{s,\ph}$, as a sum of contributions arising from phonons, spins and their interaction, and aims to calculate the thermal conductivity using an appropriate Kubo formula, it will contain 9 different terms, which describe, for example, how the phonon heat currents influence spin-heat currents and vice versa. Ultimately, the bulk gap of the spin system is the reason why a simple addition law is valid.

As the phonon Hall effect  $\kappa_{xy}^\ph$ is much smaller than the contribution $\frac{d J_d}{d T}$ from the edge anomaly, see Sec.~\ref{sec:bulk},
one can expect that the measured thermal Hall effect is dominated by the edge anomaly and thus (taking the arguments of the next section into account) approximately quantized. We would like to repeat that the coupling of edge modes and phonons was absolutely essential to derive this result, see Sec.~\ref{nonquantizedSec}. The phonon coupling enforces the temperature gradient which give rise to the pumping of energy from the edge into the bulk of the system.

According to Eq.~\eqref{jperp} there is a dissipative phonon heat current ({\it i.e.} an energy current in the direction of a temperature gradient) flowing in the perpendicular direction. For energy-conserving boundary conditions such currents usually ({\it i.e.} in the absence of edge anomalies) do not occur in the transverse direction. Here, however it is important to stress that the entropy production proportional to $j_\perp^\ph \nabla_y (1/T)$ is exactly canceled by the reduction of the current in parallel direction arising from the edge mode, which contributes proportional to   $-\frac{d J_d}{d T} \nabla_y T \nabla_x (1/T)$ to the entropy production.

\section{Edge anomaly in the presence of phonon coupling}\label{edgeAnomaly}

Next, we investigate how the equilibrium edge current in the absence of phonons, $J_e(T)$, is related to the equilibrium edge current in the presence of phonons, $J_d(T)$. As the coupling of the edge modes to the phonons immediately destroys the ballistic transport of non-equilibrium edge currents by allowing leakage into the bulk, one may worry that it also modifies the equilibrium edge current. This is directly related to the question of whether the presence of gapless phonons modifies the gravitational anomaly in the bulk and on the edge.  We will first discuss a non-perturbative argument and then check some of its assumptions by a perturbative calculation in Appendix~\ref{appendix:pert}.

Our goal will be to investigate how $J_d(T)=\langle \hat J_d \rangle$ changes when the edge Hamiltonian is changed. We describe the change by a single parameter $\lambda_e$. Changing $\lambda_e$, e.g., from $0$ to $1$ may change the band curvature, induce a coupling among edge channels, or switch on interactions and disorder along the edge (to simplify the calculation we do, however, assume translational invariance below). Most importantly, we assume that for $\lambda_e=0$ there is no coupling to the phonons while for $\lambda_e>0$ the coupling to acoustic bulk-phonons is activated. Denoting the equilibrium density matrix by $\rho$, two contribution may affect~$J_d$
\begin{eqnarray}
\frac{d}{d \lambda_e} J_d(T) = \tr\left[\frac{\partial \hat J_d}{\partial \lambda_e} \rho\right]+\tr\left[\hat J_d \frac{\partial \rho}{\partial \lambda_e} \right] \label{dJddL}
\end{eqnarray}
The second term can be identified with a standard retarded correlation function, well known from linear response theory
\begin{eqnarray}
\tr\left[\hat J_d \frac{\partial \rho}{\partial \lambda_e} \right]&=&\lim_{q_x \to 0} \lim_{\omega \to 0} \chi^{\ }_{\lambda_e}(q_x,\omega)\\
\chi^{\ }_{\lambda_e}(q_x,\omega)&=&-\frac{i}{\hbar L_x}
\int_0^\infty dt 
e^{i (\omega+ i \epsilon) t}  \times \nonumber \\ &&
\left\langle 
\left[    j_d(q_x,t) , \frac{\partial e_d(-q_x,0)}{\partial {\lambda_e}} \nonumber
\right] 
\right\rangle
\end{eqnarray}

with $L_x$ the length of the system and an infinitesimal $\epsilon>0$. 
As we calculate the change of a thermal equilibrium expectation value, the suscpetibility has to be evaluated in the limit where one takes first $\omega \to 0$ and then $q_x \to 0$ (in contrast, d.c. conductivities are calculated using the opposite order of limits).
Note that we introduced the Fourier tranformation to momentum space (but not frequency space) for the energy current density and the ${\lambda_e}$ derivative of the energy density evaluated in a strip of width $d$ along the edge. Using the Fourier transformed version of Eq.~\eqref{continuity}, $i q_x j_d(q_x,t)=-\frac{d e_d(q_x,t)}{d t}+j^\perp_{\ph,d}(q_x,t)$, we split $\chi^{\ }_{\lambda_e}$ into two terms, $\chi^{\ }_{\lambda_e}=\chi^\|_{{\lambda_e}}+\chi^\perp_{{\lambda_e}}$, where the first one is given by
\begin{eqnarray}
\chi^\|_{\lambda_e}(q_x,0)&=& \frac{1}{\hbar L_x q_x} \int_0^\infty \! dt e^{- \epsilon t} 
\left\langle 
\left[   \frac{d e_d(q_x,t)}{d t} , \frac{\partial e_d(-q_x,0)}{\partial {\lambda_e}} \nonumber
\right] 
\right\rangle \nonumber \\
&=& -\frac{1}{\hbar L_x q_x} 
\left\langle 
\left[    e_d(q_x,0) , \frac{\partial e_d(-q_x,0)}{\partial {\lambda_e}} \nonumber
\right] 
\right\rangle   \\
&=& -\int dx \, dx' \frac{e^{i q_x (x-x')} }{\hbar L_x q_x} 
\left\langle 
\left[    e_d(x) , \frac{\partial e_d(x')}{\partial {\lambda_e}} \nonumber
\right] 
\right\rangle \nonumber 
\end{eqnarray}
where we use the Fourier transformation back to position space.
In the limit $q_x \to 0$, $\chi^\|_{\lambda_e}$ evaluates to 
\begin{widetext}
\begin{eqnarray}
\chi^\|_{\lambda_e}(q_x\to0,0)&=& -\frac{1}{\hbar L_x}
\int dx \, dx' \, i 
\left\langle 
\left[  x  e_d(x) , \frac{\partial e_d(x')}{\partial {\lambda_e}} \nonumber
\right] +
\left[  x  \frac{\partial e_d(x)}{\partial {\lambda_e}}, e_d(x')  
\right] 
\right\rangle =-\left\langle \frac{\partial J_d}{\partial {\lambda_e}}
\right\rangle
\end{eqnarray}
\end{widetext}
where we used that the heat current along the edge can be written  (using the continuitiy equation) as $L_x J_d = \int dx\,dx'\, i \left\langle 
\left[  x  e_d(x) , e_d(x')  
\right] 
\right\rangle $. The contribution from $\chi_{\lambda_e}^\|$ therefore cancels exactly the first term in Eq.~\eqref{dJddL} and the only remaining term is
\begin{eqnarray}\label{chiperp}
\frac{d J_d}{d {\lambda_e}} &=& \lim_{q_x \to 0} \lim_{\omega \to 0} \chi_{\lambda_e}^{\perp}(q_x,\omega)\\
\chi_{\lambda_e}^{\perp}(q_x,\omega)&=& \frac{1}{L_x\hbar}
\int_0^\infty dt \frac{e^{i (\omega+i \epsilon) t} }{q_x} \times 
\nonumber \\ &&
\left\langle 
\left[   j^\perp_{\ph,d}(q_x,t), \frac{\partial e_d(-q_x,0)}{\partial {\lambda_e}} \nonumber
\right] 
\right\rangle \nonumber \\ 
&=&0 \quad \text{for } d\to \infty \nonumber
\end{eqnarray}
We have checked in a perturbative calculation to order ${\lambda_e}^2$ that 
$\lim_{q_x \to 0} \lim_{\omega \to 0} \chi_{\lambda_e}^{\perp}(q_x,\omega) \sim \frac{{\lambda_e}^2}{d^2}$ in two dimensions, see Appendix~\ref{appendix:pert}. The fact that $\chi_{\lambda_e}^{\perp}$ vanishes for large $d$ is plausible because by definition ${\lambda_e}$ affects only the region close to the edge while $j_\perp$ is measured
at a distance $d$ from the edge. Due to the gapless phonon modes one obtains a powerlaw decay with $d$.

For sufficiently large $d$, we can therefore conclude that $d J_d/d {\lambda_e}=0$: the coupling of the edge modes to the gapless phonon modes in the bulk has no effect on the edge anomaly in the thermodynamic limit which therefore  the first term in Eq.~\eqref{kappaAll}
 remains quantized.

\section{Contribution of Bulk Phonons to Thermal Hall Conductivity}
\label{sec:bulk}

In the previous section we have shown that the edge contribution, $d J_d/d T$ to the thermal Hall response,
 Eq.~\eqref{kappaAll}, is not affected by the coupling to phonons. To complete our discussion we have to estimate
the contribution  $\kappa^\ph_{xy}$ of the bulk phonons to the thermal Hall effect.
We ignore a tiny contribution arising for the direct coupling of the ions to an external magnetic field and study instead a -- much larger -- Berry phase correction (an `anomalous' thermal Hall effect) arisng due to the coupling of the phonons to the gapped bulk modes of the magnetic electronic or spin system.

For definiteness, we consider $2D$ phonons in a $2D$ Kitaev honeycomb model, but we expect that the main qualitative features of our result will be independent of the concrete model and generic for a large class of system showing approximately quantized thermal Hall effect.
We will derive an effective low-energy theory for the phonons which takes into account the coupling to the Majorana degrees of freedom, and then calculate the thermal Hall conductivity of this purely phononic theory.

Our starting point is the effective low-energy Hamiltonian, $H=H_K+H_p+H_{Kp}$, consisting of the standard $2D$ Kitaev model~\cite{Kitaev_2006} $H_K$, in the presence of an external magnetic field, the phonon Hamiltonian $H_p$ and the coupling of phonons to the Majorana spins. We assume that the coupling arises from the fact that the Kitaev coupling $J$ depends on the distance between atoms. The phonon-Majorana coupling constant is therefore given by  $\lambda=\frac{d J}{d r}$ and we obtain
\begin{equation}
	\mathcal{H}_I = -\lambda \sum_{{\bf n},j}
	\hat{{\bf M}}_j \cdot ({\bf u}_{{\bf n}}-
	{\bf u}_{{\bf n}+{\bf M}_j})
	i b_{{\bf n}}a_{{\bf n}+{\bf M}_j},
\end{equation}
where $\hat{\vec M_j}$ $(j=1,2,3)$ are normalized vectors connecting neighboring sites, $a$ and $b$ are the Majorana fermions on the different sublattices, and ${\vec u}_n$ is the displacement of the lattice atom at site $n$. In order to derive an effective low-energy theory we expand about the Dirac points of the Kitaev model and consider only acoustical phonon modes, where we get
\begin{equation}
	\mathcal{H} = \mathcal{H}_p + \hbar 
	v_M\int\! d^2r \, \psi^\dagger \left(
	\left[ -i \vec \nabla - \vec A)\right]\cdot
	\vec \sigma + b \, \sigma_z \right) \psi,
\end{equation}
where $\psi$ is a $2$ component fermionic spinor (arising from combining two Majorana nodes to a single Dirac node), $b$ a term due to the external magnetic field which drives the Majorana system into the quantum-Hall phase, $\mathcal{H}_p$ is the free phononic Hamiltonian and ${\bf A}$ is the dynamic field phonon field that couple to the Majoranas, given by
\begin{equation}
	{\bf A}({\bf r}) =\frac{\lambda a}{\hbar v_M}
	  \left( \begin{array}{l}
	 -\partial_x u_x({\bf r})+\partial_y u_y({\bf r})\\
	 \partial_y u_x({\bf r})+\partial_x u_y({\bf r}) 
	 \end{array}
	 \right)
	\label{eq:A_xy}
\end{equation}
where $a$ is the distance of nearest neighbors and $\vec A$ acts like a vector potential, reflecting the well-known fact~\cite{Rachel_2016} that lattice distortions can act as a `synthetic' orbital magnetic field. In principle there can also be further coupling terms, including a term $b \sigma_z (\partial_x u_x+\partial_y u_y)$ which will only renormalize the stiffness of the phonon lattice not contributing to $\kappa^{ph}_{xy}$ to leading order.

Our next task is to integrate out the Majorana degrees of freedom, and obtain an effective low-energy theory only for the phonons. Here we can use that $\vec A$ couples to the fermionic current $\vec{j}$. It is therefore possible to compute the effective action of the phonons from the conductivity tensor
$\sigma_{ij}(\omega)=\frac{\langle j_i;j_j \rangle}{i \omega}$ of the fermionic quasiparticles obtained in the usual way from the current-current correlation function (while there is no charge conservation in the parent model, we can define a conductivity for the effective low-energy model). In the gapped quantum-Hall phase and for temperatures well below the gap, $\sigma_{xx}$ vanishes but $\sigma_{xy}$ obtains the quantized value of half a conductance quantum 
\begin{eqnarray}
\sigma_{xy}=\lim_{\omega \to 0} \frac{\langle j_x;j_y \rangle}{i \omega} = \text{sign}(b) \frac{e^2}{4 \pi \hbar}
\end{eqnarray}
 
Integrating out the Majorana modes for small momenta $\vec q$ and frequencies $\omega$ results in a term in the Lagrangian of the form
$\sum_{\vec q,\omega} A_x(-\omega,-\vec q) \, \sigma_{xy}\, i \omega \, A_y(\omega,\vec q)$. Using $\hbar\partial_t A_y \approx i [H_p^0,A_y]$ we obtain after some algebra the effective Hamiltonian for the acoustic phonons in $2D$
\begin{align}
\mathcal H_p &=\mathcal H_p^0+\mathcal{H}_{p}^H  \notag \\
	\mathcal{H}_{p} &= 
\int\! \frac{d^2k}{(2 \pi)^2} \, \frac{1}{2 \mu} \, {\vec \pi}_{\bf k} {\vec \pi}_{-{\bf k}} + \frac{\mu}{2}
	{\bf u}_{\bf k} \overleftrightarrow{M}_{\vec k} {\bf u}_{-{\bf k}}
		\label{eq:H_eff_phonon} \\
	\mathcal{H}_{p}^H &= 
\frac{\lambda^2 a^2 \text{sign}(b)}{8 \pi\hbar v_M^2 \mu}\int\! \frac{d^2k}{(2 \pi)^2} (k_x^2+k_y^2) \left(u^y_{\vec k} \pi^x_{-\vec k} -u^x_{\vec k} \pi^y_{-\vec k}\right) \notag
\end{align}
where $\vec \pi$ is the momentum density of the ions, $\mu$ the 2d mass density of the solid and $\overleftrightarrow{M}$ parametrizes the strain energy
\begin{equation}
	\overleftrightarrow{M}({\bf k})= \left[\begin{array}{cc}
	u_1^2 (k_x^2+k_y^2)+u_2^2 k_x^2 & u_2^2 k_x k_y \\
	u_2^2 k_x k_y & u_1^2 (k_x^2+k_y^2)+u_2^2 k_y^2
	\end{array}\right].
\end{equation}
Due to the high symmetry of the honeycomb lattice, two parameters $u_1$ and $u_2$ are sufficient to parameterize the strain for $\vec k \to 0$. Here $u_\perp=u_1$ is the sound velocity of the transvers acoustic phonon while the longitudinal acoustic phonon obtains the velocity $u_\|=\sqrt{u_1^2+u_2^2}$ for all directions of $\hat k$.

From the Hamiltonian of Eq.~(\ref{eq:H_eff_phonon}) one can derive the equations of motions for $\vec{u}_{\vec k}$ and $\vec{\pi}_{\vec k}$, which can be written into a matrix form ${\dot x}_{\vec k} = -i \tilde{H}({\vec k}) x_{\vec k}$ with $x_{\vec k} = (\pi^x_{\vec k} \; \pi^y_{\vec k} \; u^x_{\vec k} \; u^y_{\vec k})^T$
and 
\begin{equation}
	\tilde{H}({\vec k}) = \frac{i}{2\mu}\left(\begin{array}{cc}
		\Lambda({\vec k}) & -\mu^2\overleftrightarrow{M}({\vec k}) \\
		I_{2} & \Lambda({\vec k})
	\end{array}
	\right)
\end{equation}
Here, $I_{2}$ is the $2\times 2$ identity matrix and
\begin{equation}
	\Lambda({\vec k}) = \frac{\lambda^2 a^2 \text{sign}(b)}
	{4 \pi\hbar v_M^2 }(k_x^2+k_y^2)
	\left( \begin{array}{cc} 0 & 1 \\ -1 & 0 \end{array}\right)
\end{equation}

This formulation allows us to calculate directly the phonon thermal Hall conductivity $\kappa^{\rm ph}_{xy}$~\cite{Qin_2012,Qin_2011,Zhang_2010,Wang_2009}. To do so, we find the left and right eigenvectors of  $\tilde{H}({\vec k})$, labeling them $\xi^{L/R}_{{\vec k},\sigma}$, with the corresponding eigenvalues $\omega_{{\vec k},\sigma}$. Note that the four eigenvalues come in pairs $\omega_{{\vec k},\sigma}=-\omega_{-{\vec k},-\sigma}$ which are the frequencies of the corresponding creation and annihilation operators that diagonalize the Hamiltonian. The phonon Berry curvature is then given by \cite{Qin_2012}
\begin{equation}
	\Omega^z_{{\vec k},\sigma} =
	-{\rm Im}\left({\vec \nabla}_{k}\xi^L_{{\vec k},\sigma}\times
	{\vec \nabla}_{k}\xi^R_{{\vec k},\sigma}\right)_z
\end{equation}
and $\kappa^{\rm ph}_{xy}$ is determined by $\Omega^z_{{\vec k},\sigma}$ and $\omega_{{\vec k},\sigma}$ by~\cite{Qin_2012}
\begin{equation}\label{kappaQin}
	\kappa^{\rm ph}_{xy} = -\frac{1}{\hbar T}
	\int_0^{\infty}\! d\epsilon \epsilon^2 
	\sigma^{\rm ph}_{xy}(\epsilon) 
	\frac{dn(\epsilon)}{d\epsilon}
\end{equation}
where
\begin{equation}\label{sigmaQin}
	\sigma^{\rm ph}_{xy}(\epsilon) = -\sum_{\sigma>0}
	\int\! \frac{d^2k}{(2\pi)^2}
	\Omega^z_{{\vec k},\sigma}
	\Theta(\epsilon-\hbar \omega_{{\vec k},\sigma})
\end{equation}
with the sum going only over the bands with positive frequencies.

To order $\lambda^2$ the phonon frequencies $\omega_{{\vec k},(1,2)} = u_{(\|,\perp)} k + O(\lambda^4 k^3)$ 
are not affected by the Majorana-phonon coupling. The coupling does, however, induce the Berry curvature
\begin{eqnarray}
	\Omega^z_{{\vec k},1} &=& \frac{3u_\|^2+u_\perp^2}
							{u_\|^2-u_\perp^2}
			\frac{\lambda^2 a^2 \text{sign}(b)}{8 \pi\hbar v_M^2 \mu}
			\frac{1}{u_\| k}+O(\lambda^4) \nonumber \\ 
	\Omega^z_{{\vec k},2} &=& -\frac{u_\|^2+3 u_\perp^2}
							{u_\|^2-u_\perp^2}
			\frac{\lambda^2 a^2 \text{sign}(b)}{8 \pi\hbar v_M^2 \mu}
			\frac{1}{u_{\perp} k} + O(\lambda^4)
\end{eqnarray}
of the longitudinal and transversal phonons, respectively. Note that the Berry curvature displays a $1/k$ divergence for $k \to 0$. This divergence arises because the Berry phase originate from a coupling of longitudinal and transversal phonons whose energy difference vanishes linear in $k$. The corresponding Berry flux is, however, finite. In contrast to, e.g., Berry curvatures characterizing Dirac nodes, it is not quantized and
 proportional to $\lambda^2$. The Berry phase nominally also diverges when the two sound velocities approach each other,  $u_\| \to u_\perp$ or, equivalently, for $u_2 \to 0$. Due to the opposite sign of the Berry curvature of the two band, this divergence does, however, cancel when the heat conductivity is calculated.

Using Eq.~\eqref{kappaQin} we obtain for two-dimensional phonons
\begin{align}
	\frac{\kappa^{\rm ph, 2d}_{xy}}{T} &= \frac{k_B^2}{\hbar}
	\frac{u_\|^2+u_\perp^2}{u_\|^2u_\perp^2}
	\frac{3\zeta(3)\lambda^2 a^2 \text{sign}(b)}
	{8 \pi^2 v_M^2\hbar \mu} k_B T + O(\lambda^4) \\
	& \sim 
	\frac{k_B^2}{\hbar} 
	\text{sign}(b) \left(\frac{a}{J} \frac{d J}{d r}\right)^2
	\frac{k_B T}{K_{2d} a^2}
\end{align}
with $\zeta(x)$ the Riemann zeta function, $\zeta(3)\approx 1.2$. For the last line we used $\hbar v_M \sim J a$ and identified $\mu u^2$ with the 2d bulk modulus $K_{2d}$ with units of force per length. 

For three dimensional phonons in a model of stacked Kitaev layers, the main difference is that the momentum integral in Eq.~\eqref{sigmaQin}
becomes three-dimensional. For comparison with the quantized heat conductivity, it is useful to consider the thermal conductivity per layer, obtained by multiplying the 3d heat conductivity
with the layer distance $d_z$ 
\begin{equation}\label{phonon3Dcorr}
	\frac{d_z\, \kappa^{\rm ph, 3d}_{xy} }{T}  \frac{k_B^2}{\hbar}
	\sim \text{sign}(b) \left(\frac{a}{J} 
	\frac{d J}{d r}\right)^2 \frac{k_B T^2}{\Theta_D K a^3}
\end{equation}
Here $\Theta_D$ is the Debye temperature and $K$ is the 3D bulk modulus. Assuming, e.g., a  Debye temperature of $200$\,K, a bulk modulus $K$ of $100$\,GPa and a lattice constant of $2$\,\AA, a temperature of $5$\,K, and $\frac{d J}{d r} \sim 10 \frac{J}{a}$, we obtain a phonon correction of the order of only $10^{-4}$.

The fact that phonon Berry curvatures give rise to $\kappa_{xy} \propto T^3$ has been shown previously by Qin, Zhou and Chi \cite{Qin_2012}. In their case, however, the Berry curvature and therefore $\kappa_{xy}^\ph$ was proportional to the magnetization $M$ of the system. In contrast, we find an anomalous contribution which just depends on the sign of the magnetization (or, equivalently, of $b$), not on its absolute value. This can be traced back to the quantized $\sigma_{xy}$ of the bulk Majorana arising for infinitesimal external magnetic field.
We expect that this feature is independent of the precise microscopic model and expect that an $M$ or $B$ independent $T^2$ term in $\kappa_{xy}/T$ is characteristic for a large class of chiral phases.

\section{Discussion and Conclusion} 
In our paper we put forward a straightforward interpretation of the approximately quantized Hall effect observed very recently in a chiral spin liquid state of $\alpha-$RuCl$_3$:
In the presence of an energy current, the equilibration of the edge channel with the phonons induces a temperature gradient in the edge mode. This temperature gradient implies that there is a gradient in the  steady-state heat current $J_d$ flowing at the edge. This gradient is given by $\frac{d J_d}{d T} \nabla T$ and its value is fixed by the gravitational anomaly of the edge mode. Energy conservation enforces that the energy current $j_{\perp,d}^\ph=\frac{d J_d}{d T} \nabla T$, Eq.~\eqref{continuity}, is injected into the phonon system in perpendicular direction. This leads to an approximately quantized thermal Hall effect which can be measured by tracking temperature gradients in the phonon system only.

\begin{figure}[t]
\includegraphics[width=0.9 \linewidth]{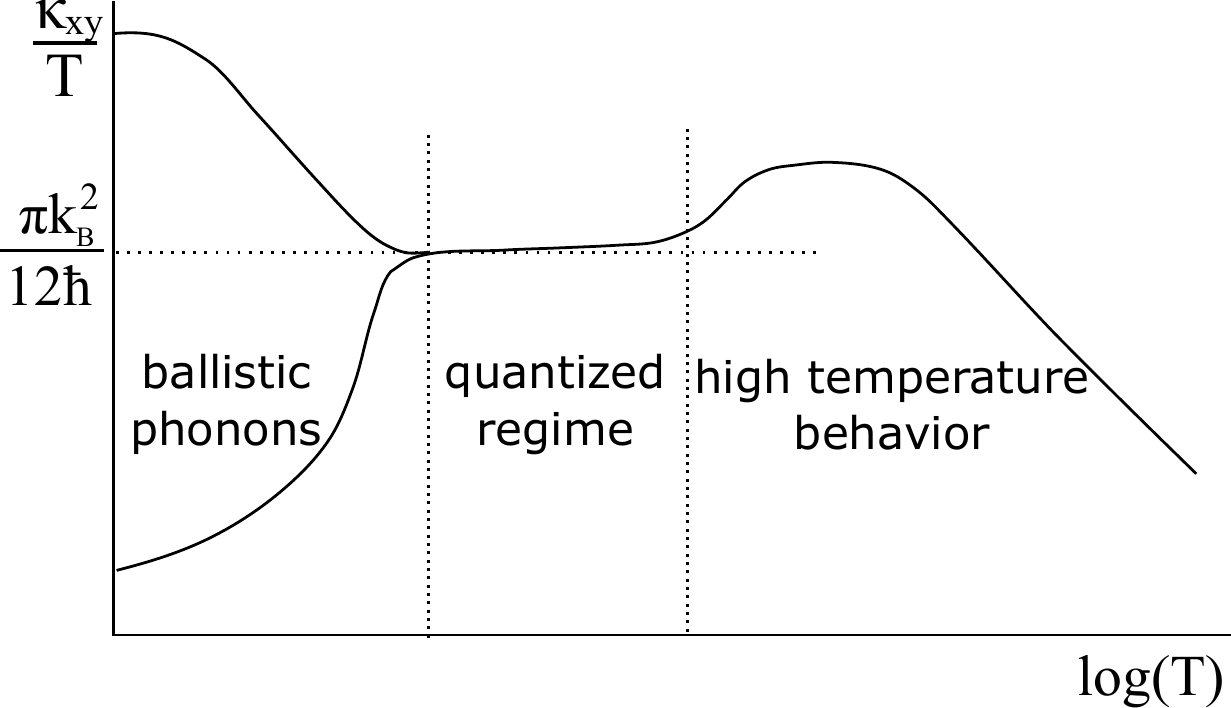}
\caption{Schematic plot of the $T$ dependence of $\kappa_{xy}$ per layer of a chiral (spin) liquid coupled to phonons. An approximately quantized Hall effect, $\frac{\kappa_{xy}}{T} =\frac{(c_r-c_l) \pi}{6 \hbar k_B}+O(T^2)$, is observed for $T$ much smaller than the bulk gap $\Delta_b$ of the chiral liquid under the condition that the phonon-phonon and the phonon-edge scattering lengthis smaller than the sample size, $\ell_\ph^\ph, \ell_\ph^e \ll L$. The leading correction for clean systems arises from an intrinsic anomalous Hall effect of phonons,
Eq.~\eqref{phonon3Dcorr}. At lower $T$ spins and phonons decouple (mean-free path larger than system size) and one obtains a non-universal result, see Sec.~\ref{nonquantizedSec}, which can be larger or smaller than the quantized value depending on contact properties. For $T \gg \Delta_b$ the thermal Hall effects both due to spins and phonons is expected to decay rapidly}
\label{fig:schematicTdependence}
\end{figure}
We have carefully checked that the amplitude of the edge anomaly (in an infinite system) is not affected by the coupling of the edge modes to the phonons. It remains topologically protected despite of the gapless phonon modes in the bulk. The only correction arises from the thermal Hall effect of the bulk phonons. 
At least in a sufficiently clean system, we expect that the leading correction to the quantized thermal Hall effect arises from Berry phases imprinted on the acoustic phonons by the chiral spin liquid. In three dimension
this leads to a correction to $\kappa_{xy}/T$ proportional to $T^2$ with a small prefactor, Eq.~\eqref{phonon3Dcorr}. Remarkably, this correction is within our approximations  independent of the size of the exteral magnetic field $B$ or the magnetization of the system but depends on the sign of $B$. 

When discussing the temperature dependence of $\kappa_{xy}$ one has to take into account that at the lowest temperatures phonons become ballistic and stop to scatter from each other and the boundary. In this regime the approximate quantization of $\kappa_{xy}$ is lost. Depending on the type of contacts used to inject the heat current and to measure the temperature it can happen, that one observes an apparent $\kappa_{xy}/T$ which is either much larger or much smaller than the quantized value, see Sec.~\ref{nonquantizedSec}.  In Fig.~\ref{fig:schematicTdependence} we show schematically the expected $T$ dependence of $\kappa_{xy}/T$. The condition for observing the quantized thermal Hall effect is that the temperature is well below the bulk gap $\Delta_b$ of the chiral liquid and that the scattering length of phonons with the edge and with themselves is smaller than the system size $L$. For $\alpha-$RuCl$_3$ we expect that the phonon correction in the relevant temperature regime is tiny, on the permille level or below. In this material  a pronounced maximum in $\kappa_{xy}$, larger than the quantized value, is observed for $T \sim \Delta_b$. The origin of this peak is an interesting open question, a possible candidate is the coupling of chiral spin fluctuations to the phonons.
For $T\gg \Delta_b$ one expects that both the phonon and spin contribution to $\kappa_{xy}$ drops rapidly.

The observation of a quantized thermal Hall effect is  the most direct measurement proving the existence of a gapped chiral (spin) liquid. Phonons help rather than hinder the observation of this effect. It is an interesting question how this physics changes  close to the quantum critical points where, e.g. as function of a magnetic field, the chiral liquid is destroyed and phonons scatter predominately from chiral critical fluctuations.

{\em Note added:} While finishing the manuscript a preprint by Ye, Halász, Savary, and Balents  was published on arXiv \cite{Ye_2018} which also emphasizes the importance of phonon coupling for the observation of a quantized thermal Hall effect by investigating the hydrodynamic equations in a finite system.

\acknowledgements
We acknowledge useful discussion with Roser Valent\'i, Yuichi Kasahara, Paul van Loosdrecht, Yuval Oreg, and Ady Stern. Financial support by the DFG (project C02 of CRC1238 and project A01 of CRC/TR183) is gratefully acknowledged.

\appendix
\section{Perturbative Analysis of Phonon Coupling to the Chiral Mode}
\label{appendix:pert}
To corroborate the picture presented in Sec.~\ref{edgeAnomaly} we analyze perturbatively the effect of the phonon coupling on the heat current. We will present at the beginning a strictly $1D$ analysis and then, based on it, an analysis of the case where the $1D$ Majorana edge mode is coupled to $2D$ phonons, which will emphasize the role of $\chi_\lambda^{\perp}$ in Eq.~(\ref{chiperp}) and its dependence on $d$, the distance from the edge.

\subsection{$1D$ case}
In $1D$  where $\chi_\lambda^\perp=0$, Eq.~(\ref{chiperp}) predicts that turning on the phonon coupling has no effect on the heat current. We will first check this statement pertubatively. To this end, we will calculate explicitly the different contributions to the heat current quadratic in the phonon coupling and show that their integrals cancel.

We consider a general $1D$ interacting Hamiltonian consisting of chiral fermions coupled locally to a single phonon band, described by $\mathcal{H} =\int\! dx \, h(x) = \int\! dx \left[ h_{0}(x) + h_{\lambda}(x)\right]$ with
\begin{align}
	h_{0}(x) &= \int\! \frac{dk dk '}{2\pi}e^{-i(k-k')x}
		\epsilon_{\frac{k+k'}{2}}c^{\dagger}_k c_{k'}  \nonumber \\
	& +\int\! \frac{dq dq '}{2\pi}e^{-i(q-q')x}
		\omega_{\frac{q+q'}{2}}a^{\dagger}_q a_{q'}, \nonumber \\
	h_{\lambda}(x) &= \int\! \frac{dk dk ' dq}{2\pi}e^{-i(k-k'-q)x}
		\lambda_{k,k'}^q c^{\dagger}_k c_{k'} a_q + {\rm h.c.}
\end{align}
We make no assumptions as to fermionic energy bands $\epsilon_{k}$, bosonic energy bands $\omega_{q}$ or the coupling coefficients $\lambda_{k,k'}^q$.

We proceed to derive the heat current operator from the $1D$ continuity equation $\partial_x j(x) + \partial_t h(x) =0$ in momentum space
\begin{equation}
	j(K) = -\frac{1}{K}\int\! dx dx' e^{-iKx} \left[h(x),h(x')\right]
\end{equation}
and the total heat current is then $J_T = \lim_{K\to 0} j(K)/L_x,$ with $L_x=\delta_k(0)$ the size of the system. Carrying out the commutation relations and taking the limit $K\to 0$ we can divide the contributions to $j(K)$ into four parts, arising from the different components of the Hamiltonian, $\langle J_T \rangle = J_{00}+J_{0\lambda}+J_{\lambda 0}+J_{\lambda\lambda}$ with
\begin{eqnarray}
	J_{\rm 00} &=&
	\frac{1}{L_x}\int\! dk v_k \epsilon_k 
	\langle c^{\dagger}_k c_k\rangle +
	\frac{1}{L_x}\int\! dq u_q \omega_q
	\langle a^{\dagger}_q a_q\rangle, \nonumber \\
	J_{0 \lambda } &=& \frac{1}{L_x}\int\! dk dk' dq 
	\bigg\{
	\partial_k
	\left[\delta(k\!-\!k'\!-\!q)\lambda^q_{k,k'} \epsilon_k\right]+
	\nonumber \\ &&
	\partial_{k '}
	\left[\delta(k\!-\!k'\!-\!q)\lambda^q_{k,k'} \epsilon_{k '}\right]+
	\partial_q
	\left[\delta(k\!-\!k'\!-\!q)\lambda^q_{k,k'} \omega_q\right]
	\nonumber \\ &&
	-\delta(k\!-\!k'\!-\!q)\lambda_{k,k'}^q 
	\frac{v_k+v_{k'}+u_q}{2}\bigg\}\langle c^{\dagger}_k c_{k'} a_q
					\rangle
	+{\rm c.c.}
	, \nonumber \\
	J_{\lambda 0} &=& \frac{1}{L_x}
	\int\! dk dk' dq \partial_k\delta(k\!-\!k'\!-\!q)\lambda_{k,k'}^q
	\left(\omega_q+\epsilon_{k'}-\epsilon_{k}\right)
	\langle c^{\dagger}_k c_{k'} a_q \rangle\nonumber \\ &&
	 + {\rm c.c.}
	, \nonumber \\
	J_{\lambda \lambda} &=& \frac{1}{L_x}
	\int\! dk dk' dq dp
	\partial_k\delta(k\!-\!k'\!-\!q)
	\lambda_{k,k'}^q\lambda_{p+q,p}^{q\;*}
	\langle c^{\dagger}_k c_{k'}c^{\dagger}_p c_{p+q}\rangle
	\nonumber \\ &&+
	\int\! dk dk' dq dq'
	\partial_k\delta(k\!-\!k'\!-\!q)
	\lambda_{k,k'}^q\lambda_{p,k'}^{p-k'\;*}
	\langle c^{\dagger}_k c_{p}a^{\dagger}_{p-k'} a_{q}\rangle
	\nonumber \\ &&-
	\int\! dk dk' dq dq'
	\partial_k\delta(k\!-\!k'\!-\!q)
	\lambda_{k,k'}^q\lambda_{k,p}^{k-p\;*}
	\langle c^{\dagger}_p c_{k'}a^{\dagger}_{k-p} a_{q}\rangle
	\nonumber \\ && + {\rm c.c.} +
	\ldots \nonumber \\
\end{eqnarray}
with $\hbar v_k = \partial_k \epsilon_k$ and $\hbar u_q=\partial_q \omega_q$, and the $\ldots$ in the expression for $J_{\lambda \lambda}$ stands for terms containing two bosonic annihilation or creation operators, which will not contribute to the current at order $\lambda^2$.

\begin{figure}[t]
\begin{tabular}{cccc}
a) & 
\includegraphics[width=0.15\textwidth]{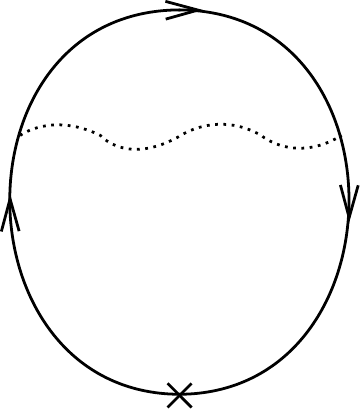} &
b) & 
\includegraphics[width=0.15\textwidth]{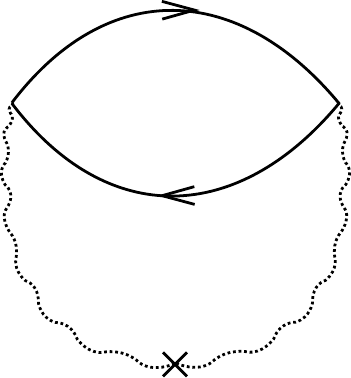}
\\ \\
c) & 
\includegraphics[width=0.15\textwidth]{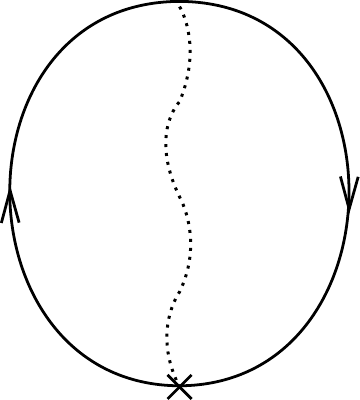} &
d) & 
\includegraphics[width=0.085\textwidth]{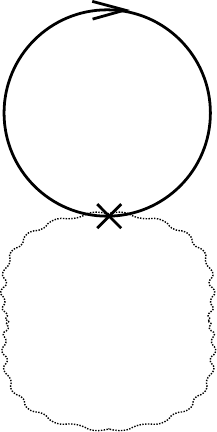}
\end{tabular}
\caption{Feynman diagrams used in calculating the expectation values of the heat current to leading order. Here, straight continuous lines are the fermions propagators and dotted wiggly lines are phonons propagators. The `x' marks the space-time point at which the expectation value is evaluated. The different diagrams are a) the $\lambda^2$ contribution to $\langle c^{\dagger}_k c_k\rangle$, b) the $\lambda^2$ contribution to $\langle a^{\dagger}_{\vec q} a_{{\vec q}'}\rangle$, c) the $\lambda$ contribution to $\langle c^{\dagger}_k c_{k '} a_{\vec q} \rangle$, d) the $\lambda^0$ contribution to $\langle c^{\dagger}_{k 
} c_{k '} a^{\dagger}_{\vec q} a_{{\vec q}'} \rangle$}
\label{fig:diagrams}
\end{figure}

We now show that the leading corrections of order $\lambda^2$ cancel when summing all the contributions. We start by focusing on $J_{00}$ and $J_{0\lambda}$, and expand the expectation values to evaluate the leading contribution in $\lambda^2$, by calculating the diagrams in Fig.~\ref{fig:diagrams}. This results in
\begin{align}
	J_{00} &= J_{00}^{0} -
	\frac{1}{L_x}
	\int\! dk dk' dq \delta^2(k\!-\!k'\!-\!q) |\lambda_{k,k'}^q|^2
	\times \\ & \big[\frac{v_k \epsilon_k}{\beta^2}
	\sum_{n,m}g^2_k(i\omega_n)g_{k'}(i\omega_m)d_q(i\omega_n-i\omega_m)
	+ \nonumber \\ &
	\frac{v_{k'}\epsilon_{k'}}{\beta^2}\sum_{n,m}g_{k}(i\omega_n)
	g^2_{k'}(i\omega_m)d_q(i\omega_n-i\omega_m)+
	\nonumber \\ &
	\frac{u_q\omega_q}{\beta^2}\sum_{n,m}g_{k}(i\omega_n)
	g_{k'}(i\omega_m)d^2_q(i\omega_n-i\omega_m)
	\big]\!+\! O(\lambda^4), \nonumber \\
	J_{0\lambda} &= \frac{2}{\hbar L_x}{\rm Re}\int\! dk dk' dq 
	\bigg\{
	\partial_k
	\left[\delta(k\!-\!k'\!-\!q)\lambda^q_{k,k'} \epsilon_k\right]+
	\nonumber \\ &
	\partial_{k '}
	\left[\delta(k\!-\!k'\!-\!q)\lambda^q_{k,k'} \epsilon_{k '}\right]+
	\partial_q
	\left[\delta(k\!-\!k'\!-\!q)\lambda^q_{k,k'} \omega_q\right]
	\nonumber \\ &
	-\delta(k\!-\!k'\!-\!q)\lambda_{k,k'}^q 
	\hbar\frac{v_k+v_{k'}+u_q}{2}\bigg\}\delta(k-k'-q)
	\lambda_{k,k'}^{q\;*}
	\times \nonumber \\ &
	\frac{1}{\beta^2}
	\sum_{n,m}g_{k}(i\omega_n) g_{k'}(i\omega_m)
	d_q(i\omega_n-i\omega_m)
	+O(\lambda^4)\label{a5}
\end{align}
with $g_k(i\omega_n) = (i\omega_n-\epsilon_k)^{-1}$, $d_q(i\Omega_n)=(i\Omega_n-\omega_q)^{-1}$ the bare Matsubara Green functions of the fermions and the phonons, $\omega_n=(2n+1)\pi/\beta$ ($\Omega_n = 2n\pi/\beta$) the Matsubara fermionic (bosonic) frequencies and $\beta$ the inverse temperature. In the expression for $J_{00}$, the term $J_{00}^0$ stands for the nonineracting $\lambda^0$ contribution, and we have changed variables $k \leftrightarrow k'$ to get the second term in the parenthesis in the integrand, with the multiplier $v_{k'} \epsilon_{k'}$. To see the cancellation between the $\lambda^2$ contribution, one first uses $$
{\rm Re}\left\{\left[2\partial_k(f_{k,k'}^q\epsilon_k)-f_{k,k'}^q\hbar v_k\right]f_{k,k'}^{q\; *}\right\} = \partial_k\left[|f_{k,k'}^q|^2 \epsilon_{k}\right]$$
with $f_{k,k'}^q=\delta(k-k'-q)\lambda_{k,k'}^q$ to reorder terms in Eq.~\eqref{a5}. Then, using integration by parts while noticing that $\partial_k g_{k} = -\hbar v_k g_{k}^2$, $\partial_q d_{q} = -\hbar u_q d_{q}^2$, the terms in $J_{0\lambda}$ attain the exact form of the terms in $J_{00}$, but with the opposite sign. Boundary terms are exponentially small in temperature and are neglected.

Next, we show that the leading contributions to $J_{\lambda\lambda}$ and to $J_{\lambda 0}$ cancel each other as well. Evaluating the expectations values to leading order, we get
\begin{eqnarray}
	J_{\lambda 0} &=& \frac{1}{\hbar L_x}
	\int\! dkdk'dq\partial_k\delta^2(k\!-\!k'\!-\!q)
	|\lambda_{k,k'}^q|^2 \times \nonumber \\ &&
	\left[f(\epsilon_{k'})+n(-\omega_q)\right]
	\left[f(\epsilon_k)-f(\epsilon_{k'}+\omega_q)\right]+O(\lambda^4),
	\nonumber \\ 
	J_{\lambda\lambda} &=& \frac{1}{\hbar L_x}
	\int\! dkdk'dq\partial_k\delta^2(k\!-\!k'\!-\!q)
	|\lambda_{k,k'}^q|^2 \times \nonumber \\ &&
	\left\{f(\epsilon_{k})\left[(1\!-\!f(\epsilon_{k'})\right]+
	\left[f(\epsilon_k)\!-\!f(\epsilon_{k'})\right]n(\omega_q)\right\}
	+O(\lambda^4) \nonumber \\
\end{eqnarray}
where $f(\epsilon)$ and $n(\omega)$ are the Fermi-Dirac and Bose-Einstein distributions, respectively. These contributions also sum to zero.

To conclude, we have shown that in $1D$ the fermion-phonon interaction does not affect the heat current to order $\lambda$, a result conistent with the non-perturbative proof in the main text.

\subsection{$2D$ case}
fThe generalization of the $1D$ case to a model which include $2D$ phonons is straightforward. The local free phonon Hamiltonian now reads
\begin{equation}
	h_0^\ph(\vec{r}) = \int\! \frac{d^2q d^2 q'}{(2\pi)^2}
	e^{-i(\vec{q}-\vec{q}')\vec{r}}\omega_{\frac{\vec{q}+\vec{q}'}{2}}
	a^{\dagger}_{\vec{q}} a_{\vec{q}'},
\end{equation}
and all  integrals over phonon momenta $q$ in $h_{\lambda}$ are now two-dimensional. Notice, however, that the coupling is still restricted to $y=0$ and the Majorana mode, as well as the coupling, is $1D$. While momentum is still conserved in the $x$-direction, the coupling to the edge mode at $y=0$ breaks momentum conservation in the $y$-direction. The formulation here is slightly different than in Sec.~\ref{SecThermal}, as the phonons travel on both sides of the Majorana edge mode. This is more convenient for computational reasons, and we believe that it does not affect the main features of the heat current that we are interested in.

We are interested in the total heat current in the $x$-direction, going through a region of width $d$ in the $y$ direction $J_d = \int_{-\infty}^{\infty}\! dx\int_{-d/2}^{d/2}\! dy \, j_x(\vec{r})$. In the limit $d\to\infty$ the arguments presented in the $1D$ case hold, and we can again show that all contributions of order $\lambda^2$ vanish. For finite $d$, however, heat current can leak outside the region $d$ in the perpendicular direction. It will be carried only by phonons, and we can evaluate it as
\begin{equation}
	\Delta J_x^{\ph}(d) = 
	J_x^{\ph,{\rm inf}} - \int\! dx \int_{-d/2}^{d/2}
	dy \, j^{\ph}(\vec{r})
\end{equation}
where $\vec{j}^{\ph}$ is the heat current carried by the phonons, and $J_x^{\ph,{\rm inf}}=\int\!d^2r\, j_x^{\ph}({\vec r})$.

\begin{figure}[b]
\begin{tabular}{cc}
a) & 
\includegraphics[width=0.4\textwidth]{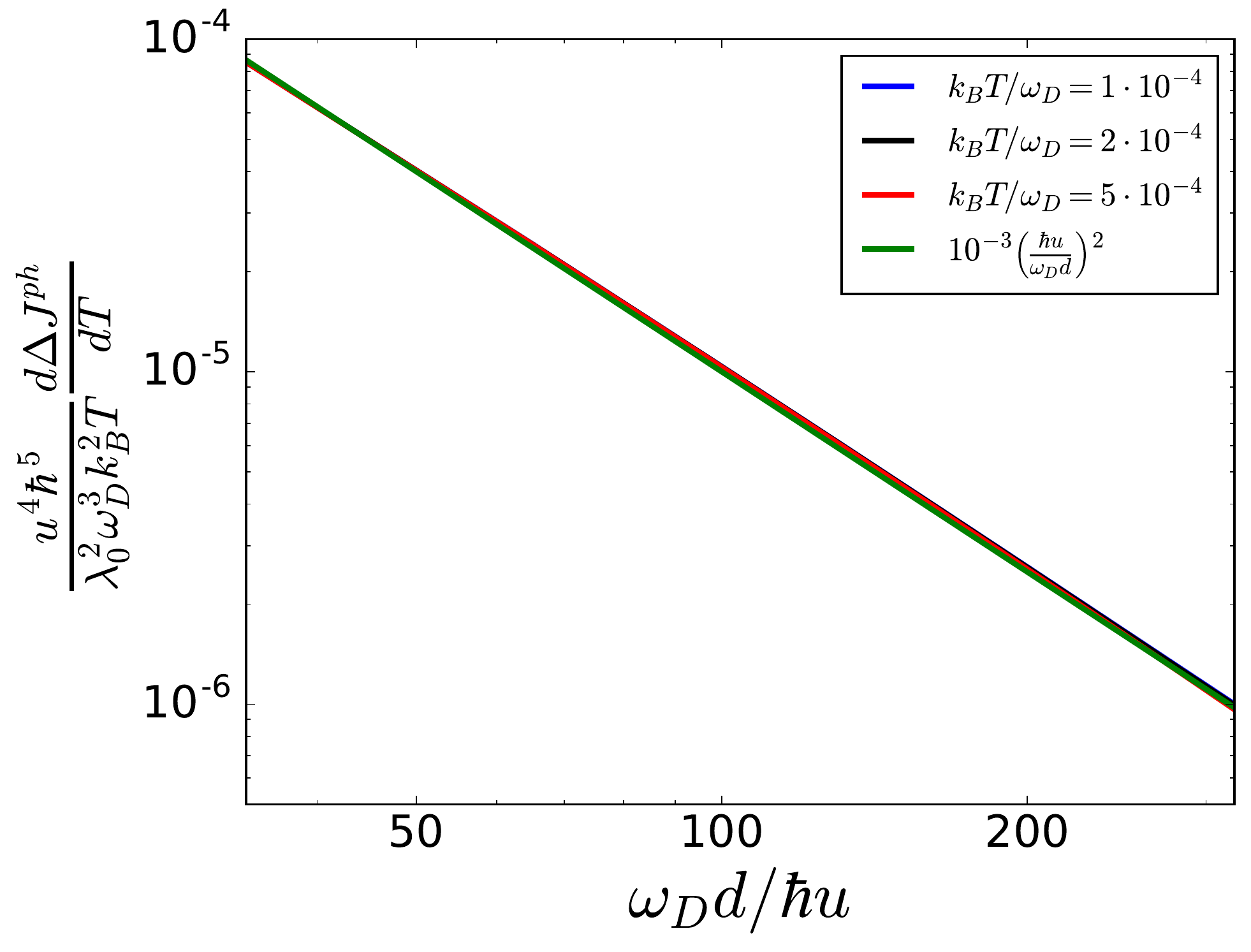} \\ \\
b) & \includegraphics[width=0.4\textwidth]{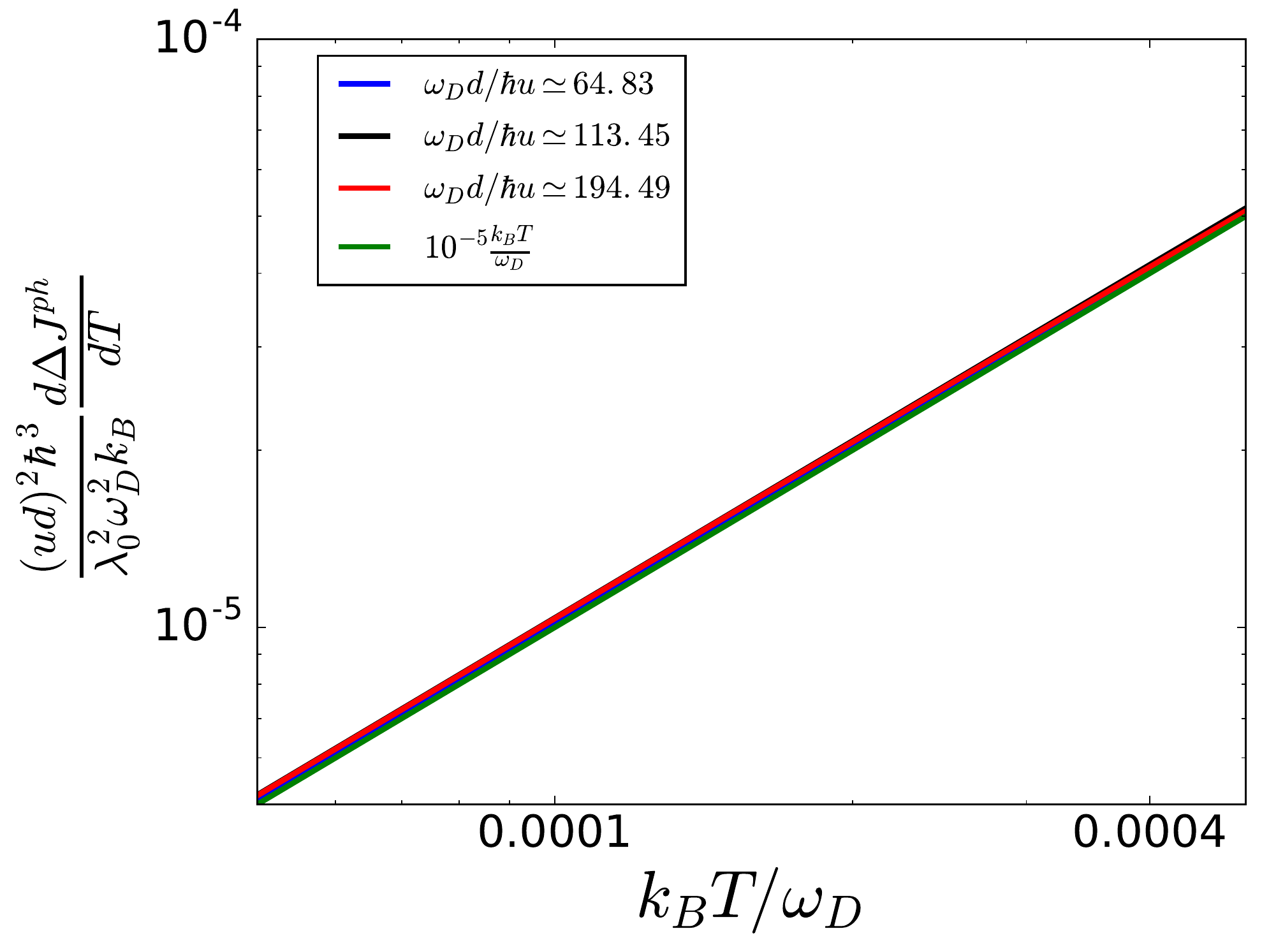}
\end{tabular}
\caption{(color online) Dependence of the heat current flowing in the perpendicular direction on a) the distance $d$ from the edge and b) the temperature. Here we plot the temperature derivative of the difference between the heat current for infinite system and the heat current when considering only a slab of width $d$, for different temperatures in (a) and different distances in (b). The green line shows a) $1/d^2$ behavior, and b) linear $T$ behavior.}
\label{fig:d_dependence}
\end{figure}

We write $\int\! dx \,j_x^{\ph}(\vec{r})$ to leading order in the Majorana-phonn interaction explicitly as
\begin{widetext}
\begin{eqnarray}
	\int\! dx j_x^{\ph}(\vec{r}) &=&
	\int\! \frac{dk dk' dq_y dq_y'}{2\pi}e^{-i(q_y-q_y')y}
	u^x_{(k-k',q_y)} \omega_{(k-k',q_y)}
	\lambda_{k,k'}^{(k-k',q_y)}\lambda_{k',k}^{(k-k',q_y') \;*}
	\left[ f(\epsilon_{k'})-f(\epsilon{k}) \right]	
	\times \nonumber \\ &&
	\frac{
	\left(\omega_{(k-k',q_y')}\!-\!\epsilon_k\!+\!\epsilon_{k'}\right)
	n(\omega_{(k-k',q_y)})\!-\!
	\left(\omega_{(k-k',q_y)}\!-\!\epsilon_k\!+\!\epsilon_{k'}\right)
	n(\omega_{(k-k',q_y')})\!+\!
	\left(\omega_{(k-k',q_y)}\!-\!\omega_{(k-k',q_y')}\right)
	n(\epsilon_k-\epsilon_{k'})
	}
	{(\omega_{(k-k',q_y)}-\omega_{(k-k',q_y')})
	(\omega_{(k-k',q_y)}-\epsilon_k+\epsilon_{k'})
	(\omega_{(k-k',q_y)}-\epsilon_k+\epsilon_{k'})
	}\nonumber \\ 
\end{eqnarray}
\end{widetext}
Note that the heat current is proportional to $\lambda^2$, as the bare heat current of the phonons is zero due to the fact that the energy band of the phonons is an even function of the momentum $\omega_{\vec{q}}=\omega_{-\vec{q}}$ which makes $\vec{u}_{\vec{q}}\omega_{\vec{q}}n(\omega_{\vec{q}})$ an odd function whose integral vanishes.

To evaluate $\Delta J_x^{\ph}(d)$ numerically we consider a specific model, namely such of a Majorana edge state with dispersion $\epsilon_k=v_M k$ coupled locally to a single band of acoustic $2D$ phonons with dispersion $\omega_{\vec q} = uq$. For coupling to electrons, the minimal coupling is to their density $h_{\lambda}(x)=\lambda \partial_x\phi(x)\psi^{\dagger}(x)\psi(x)$ where $\phi(x)$ is the phononic field. However, such a coupling is not possible for a Majorana edge, as $\psi^{\dagger}(x)\psi(x) = \psi^2(x) \sim 1$, and we must consider coupling of the form $h_{\lambda}(x)=\lambda \partial_x\phi(x)\psi^{\dagger}(x)(-i\partial_x)\psi(x)$. This results in $\lambda_{k,k'}^{\vec{q}} = \lambda_0 \hbar v_M(k+p)q_x/\sqrt{\omega_{\vec{q}}}\exp(-\omega_{\vec{q}}/\omega_D)$, where $\lambda_0$ is a coupling constant with dimensions of square root of energy times area, and we introduced the Debye energy $\omega_D$ via a soft cutoff on the coupling. Finally, for computational convenience, we slightly change the formula for $\Delta J_x^{\ph}(d)$ by replacing the hard cutoff at $\pm d/2$ with a soft Gaussian of width $d$
\begin{equation}
	\Delta J_x^{\ph}(d) \simeq
	J_x^{\ph,{\rm inf}} - 
	\frac{1}{\sqrt{\pi}}
	\int\! d^2r e^{-(y/d)^2} j^{\ph}(\vec{r})
\end{equation}
which we believe does not affect the results qualitatively, and evaluate the temperature derivative of this quantity $d \Delta J_x^{\ph}(d)/ dT$.

The numerical integration shows that the phonon correction has the following scaling behavior
\begin{equation}
	\Delta J^\ph_x(d) \sim \frac{\lambda_0^2 \omega_D (k_B T)^2}{(u d)^2 \hbar^3}
\end{equation}
and therefore vanishes for large $d$, $\Delta J^\ph_x(d)\propto 1/d^2$, see Fig.~\ref{fig:d_dependence}. This shows that at least to order $\lambda^2$ and in the thermodynamic limit the edge anomaly is not affected by the coupling to gapless bulk phonons, see main text.

\bibliography{t_h_bib}

\end{document}